\documentclass[letter]{ptptex}
\usepackage{amsmath}
\usepackage{amssymb}
\usepackage[dvipdfm]{graphicx}

\title{Automatic Hermiticity}

\author{Keiichi \textsc{Nagao}$^{1,2,}$\footnote{E-mail: 
nagao@mx.ibaraki.ac.jp, nagao@nbi.dk} 
and Holger Bech \textsc{Nielsen}$^{2,}$\footnote{E-mail: hbech@nbi.dk}}

\inst{$^{1}$Faculty of Education, Ibaraki University, Mito 310-8512, Japan \\
$^{2}$Niels Bohr Institute, University of Copenhagen, \\
17 Blegdamsvej Copenhagen $\phi$, Denmark}

\recdate{October 26, 2010; Revised January 26, 2011}

\abst{We study a diagonalizable Hamiltonian that is not at first hermitian. 
Requirement that a measurement shall not change one Hamiltonian 
eigenstate into another one with a different eigenvalue imposes that
an inner product must be defined so as to 
make the Hamiltonian normal with regard to it. 
After a long time development with the non-hermitian Hamiltonian, 
only a subspace of possible states will effectively survive. 
On this subspace the effect of the anti-hermitian part of the 
Hamiltonian is suppressed, and the Hamiltonian becomes hermitian. 
Thus hermiticity emerges automatically, and we have {\it no reason to 
maintain that at the fundamental level the Hamiltonian should be hermitian}. 
If the Hamiltonian is given in a local form, 
a conserved probability current density can be constructed 
with two kinds of wave functions. 
We also point out a possible misestimation of a past state 
by extrapolating back in time with the hermitian Hamiltonian. 
It is a seeming past state, not a true one. }

\begin{document}
\maketitle

\vspace*{0.5cm}
\noindent{\it Introduction}\hspace*{7mm}
In quantum theory the action $S$ is real and 
thought to be more fundamental than the integrand $\exp( \frac{i}{\hbar}S )$ 
of the Feynman Path Integral. 
But if we assume that the integrand is more fundamental than the action in quantum theory, 
then it is naturally thought that since the integrand is complex, 
the action also could be complex.
Based on this assumption and other related works 
in some  backward causation developments\cite{ownctl}  inspired by general relativity
and the non-locality explanation 
of fine-tuning problems \cite{nonlocal}, the complex action theory 
has been studied intensively by one of the authors (H.B.N.) and Ninomiya\cite{own}. 
Indeed, many interesting suggestions have been made for Higgs mass\cite{Nielsen:2007mj}, 
quantum mechanical philosophy\cite{newer}, 
some fine-tuning problems\cite{Nielsen2010qq,degenerate}
and black holes\cite{Nielsen2009hq}.
In Refs.~\citen{own,Nielsen:2007mj,newer,Nielsen2010qq,degenerate,Nielsen2009hq} 
they studied a future-included version, that is to say, 
the theory including not only a past time but also a future time 
as an integration interval of time. 
In contrast to the above references, in this paper we 
consider a future-not-included version.

We shall study a system defined by a non-hermitian Hamiltonian $H$, 
which is correlated to the complex action, 
and look at the time-development of some state. 
As for non-hermitian Hamiltonians, they have traditionally been used for dissipative systems. 
Especially, we note that a class of non-hermitian Hamiltonians 
satisfying the PT symmetry has been intensively studied 
in various directions\cite{PTsym_Hamiltonians,Geyer}. 
It has the strong property that the eigenvalues are real, and thus 
it has been shown that the Hamiltonians not only give completely consistent 
quantum theories but that they are also now the subject of significant experimental works. 
Such a PT symmetry has been considered also in a different context\cite{Erdem:2006zh}.

On the other hand we study a general non-hermitian Hamiltonian in this paper. 
As we know, the time development operator is non-unitary, 
and thus the probability conservation is not held. 
Also, the eigenvalues of the Hamiltonian are not real in general. 
Furthermore, since the eigenstates are not orthogonal, 
a transition that should not be possible could be measured. 
From these properties it does not look a physically reasonable theory, in contrast 
to the PT symmetric Hamiltonian formalism.  
But, contrary to our naive expectation, we shall find that 
it could be {\em effectively} a physically reasonable theory via two procedures.

The first procedure is to define a physically reasonable inner product $I_Q$ such that 
the eigenstates of the Hamiltonian get orthogonal with regard to it, and thus 
it gives us the true probability for a transition from some state to another. 
As we shall see later, $I_Q$ makes the Hamiltonian normal with regard to it. 
In other words $I_Q$ has to be defined for consistency so that the Hamiltonian 
--- even if it cannot be made hermitian --- at least be normal.  
We explain how a reasonable physical assumption about the probabilities leads to 
the proper inner product $I_Q$, 
and define a hermiticity with regard to $I_Q$, $Q$-hermiticity.

The second procedure is to use a mechanism of 
suppressing the effect of the anti-hermitian part of the Hamiltonian $H$ 
after a long time development. 
This is speculated in Ref.~\citen{originsym}. 
In this paper we shall explicitly show the mechanism 
with the help of the proper inner product $I_Q$.  
For the states with high imaginary part of eigenvalues of $H$, 
the factor $\exp\left(-\frac{i}{\hbar} H(t - t_0) \right)$ will exponentially grow with $t$ 
and faster the higher the eigenvalues are.  
After a long time the states with the highest imaginary part of eigenvalues of $H$ 
get more favored to result than others. 
That is to say, the effect of the imaginary part, which shall be shown to correspond to 
the anti-$Q$-hermitian part of $H$, gets attenuated. 
Utilizing this effect to normalize the state, 
we can effectively obtain a $Q$-hermitian Hamiltonian.

\vspace*{1mm}
\noindent
{\it Physical significance of an inner product} \hspace*{7mm}
The Born rule of quantum mechanics is well-known in the form: 
When a quantum mechanical system prepared in a state $| i \rangle$ at time $t_i$ time-develops into 
$| i (t_f) \rangle = e^{-\frac{i}{\hbar}H(t_f -t_i)}| i \rangle$ at time $t_f$, 
we will measure it in a state $|f \rangle$ with the probability 
$P_{f ~\text{from} ~i} = | \langle f | i(t_f) \rangle |^2$. 
We note that the probability depends on how we define an inner product of the Hilbert space. 
A usual inner product is defined as a sesquilinear form. 
We denote it as $I( | f \rangle, | i(t_f) \rangle ) \equiv \langle f | i(t_f) \rangle$. 
It is $|I( | f \rangle, | i(t_f) \rangle )|^2$ that we measure 
by seeing how often we get $| f \rangle $ from $| i(t_f) \rangle $. 
Measuring the transition of superposition like $c_1 | a \rangle + c_2 | b \rangle$ repeatedly, 
we can extract the whole form of $I(| f \rangle, | i(t_f) \rangle)$ of any two states 
by using the sesquilinearity.

To consider an inner product in our theory with non-hermitian Hamiltonian $H$, 
we assume that $H$ is diagonalizable, and diagonalize $H$ by using a non-unitary operator $P$ as 
\begin{equation}
H = PD P^{-1}. 
\end{equation}
We introduce an orthonormal basis $| e_i \rangle (i=1, \ldots)$ satisfying 
$\langle e_i | e_j \rangle = \delta_{ij}$ by 
$D | e_i \rangle = \lambda_i   | e_i \rangle$, 
where $\lambda_i (i=1, \ldots)$ are generally complex.
We also introduce the eigenstates $| \lambda_i \rangle$ of $H$ by 
$| \lambda_i \rangle = P | e_i \rangle$, which obeys 
\begin{equation}
H | \lambda_i \rangle = \lambda_i | \lambda_i \rangle.
\end{equation}
We note that $| \lambda_i \rangle$ are not orthogonal to each other in the usual inner product $I$, 
$\langle \lambda_i | \lambda_j \rangle \neq \delta_{ij}$.

As we are prepared, let us apply the usual inner product $I$ to our theory 
with the non-hermitian Hamiltonian $H$, 
and consider a transition from an eigenstate $| \lambda_i \rangle$ 
to another $| \lambda_j \rangle ~(i \neq j)$ fast in time $\Delta t$. 
Then, though $H$ cannot bring the system from one eigenstate $| \lambda_i \rangle$ 
to another one $| \lambda_j \rangle ~(i \neq j)$,  
the transition can be measured, that is to say, 
$|I(| \lambda_j \rangle, \exp\left( -\frac{i}{\hbar} H \Delta t \right) |\lambda_i \rangle )|^2 \neq 0$, 
since the two eigenstates are not orthogonal to each other. 
We believe that such a transition should be prohibited in a reasonable theory, 
based on the philosophy that a measurement --- even performed in a short time --- is fundamentally 
a physical development in time. 
Thus we think that the eigenstates have to be orthogonal to each other.

\vspace*{1mm}
\noindent
{\it A proper inner product and hermitian conjugate}\hspace*{7mm}
Since we are physically entitled to require that a truly functioning measurement procedure 
must necessarily have reasonable probabilistic results, we attempt to construct 
a proper inner product $I_Q( | f \rangle, | i \rangle ) \equiv \langle f |_Q ~i \rangle$ 
with the property that the eigenstates $| \lambda_i \rangle$ and $| \lambda_j \rangle$ get orthogonal to each other, 
\begin{equation}
I_Q( | \lambda_i \rangle, | \lambda_j \rangle ) = \delta_{ij}. \label{IQ_delta_ij}
\end{equation}
We believe that the true probability is given by such a proper inner product $I_Q$, 
based on which 
the Hamiltonian is conserved even if it is not hermitian and typically has complex eigenvalues. 
This condition applies to not only the eigenstates of the Hamiltonian 
but also those of any other conserved quantities. 
The transition from an eigenstate of such a conserved quantity to another eigenstate 
with a different eigenvalue should be prohibited in a reasonable theory.

Let us write the proper inner product with the property of Eq.~(\ref{IQ_delta_ij}) in the following form,  
\begin{equation}
I_Q(|\psi_2 \rangle,|\psi_1 \rangle)=\langle \psi_2 |_Q \psi_1 \rangle \equiv \langle \psi_2 | Q | \psi_1 \rangle ,
\end{equation}
where $Q$ is some operator chosen appropriately, and $| \psi_1 \rangle$ and $|\psi_2 \rangle$ are arbitrary states. 
Of course in the special case of the Hamiltonian $H$ being already hermitian $Q$ would be 
the unit operator. 
In the usual real action theory the usual inner product $I$ is defined to satisfy 
$\langle \psi_1(t) | \psi_2(t) \rangle = \langle \psi_2(t) | \psi_1(t) \rangle^*$. 
Hence we impose a similar relation on $I_Q$ as 
\begin{equation}
\langle \psi_1(t) |_Q \psi_2(t) \rangle = \langle \psi_2(t) |_Q \psi_1(t) \rangle^*. 
\end{equation} 
Then we obtain a condition $Q^\dag=Q$, namely, $Q$ has to be hermitian.

Via the inner product $I_Q$, we define the corresponding hermitian conjugate $\dag_Q$ 
for some operator $A$ by 
\begin{equation}
\langle \psi_2 |_Q A | \psi_1 \rangle^* = \langle \psi_1 |_Q {A}^{\dag_Q} | \psi_2 \rangle . \label{def_dag_n}
\end{equation}
Since the left-hand side can be expressed as 
$\langle \psi_2 |_Q {A} | \psi_1 \rangle^* = 
\langle \psi_1 | {A}^\dag Q^\dag | \psi_2 \rangle$, 
an explicit form of the $Q$-hermitian conjugate of $A$ is given by 
\begin{equation}
A^{\dag_Q} = Q^{-1} A^\dag Q.
\end{equation} 
${\dag_Q}$ is introduced for operators, but we can formally define 
$\dag_Q$ for kets and bras, too. 
We define ${\dag_Q}$ for kets and bras as 
$| \psi_1 \rangle^{\dag_Q} \equiv \langle \psi_1 |_Q $ and 
$\left(\langle \psi_2 |_Q \right)^{\dag_Q} \equiv | \psi_2 \rangle$. 
Then we can manipulate $\dag_Q$ like a usual hermitian conjugate $\dag$. 
When $A$ satisfies $A^{\dag_Q} = A$, 
we call $A$ $Q$-hermitian. 
This is the definition of the $Q$-hermiticity.  
Since this relation can be expressed as $Q A = ( Q A )^\dag$, 
when $A$ is $Q$-hermitian, $QA$ is hermitian, and vice versa. 
We note that in Ref.~\citen{Geyer} a similar inner product has been studied in more detail, 
and a criterion for identifying a unique
inner product through the choice of physical observables has also been  provided.

If some operator $A$ can be diagonalized as $A=P_A D_A P_A^{-1}$, 
then $Q$-hermitian conjugate of $A$ is expressed as  
$A^{\dag_Q} = Q^{-1} (P_A^\dag )^{-1} D_A^\dag P_A^\dag Q$. 
If we choose $Q$ as $Q= (P_A^\dag)^{-1} P_A^{-1}$, which satisfies $Q^\dag=Q$, 
we have $A^{\dag_Q} = P_A D_A^\dag P_A^{-1}$.
Therefore, if the diagonal components of $D_A$ are real, namely, $D_A^\dag=D_A$, 
then $A$ is shown to be $Q$-hermitian. 
In the following we define $Q$ by 
\begin{equation}
Q= (P^\dag)^{-1} P^{-1}
\end{equation}
with the diagonalizing matrix $P$ of the non-hermitian Hamiltonian $H$. 
We note that $I_Q$ is different from the CPT inner product defined in the PT symmetric 
Hamiltonian formalism\cite{PTsym_Hamiltonians}.

\vspace*{1mm}
\noindent
{\it $Q$-normality of the Hamiltonian}\hspace*{7mm}
To prove that the non-hermitian Hamiltonian $H$ is $Q$-normal, 
i.e. normal with regard to the inner product $I_Q$, we first define 
\begin{equation}
``P^{\dag_Q}"
\equiv
\left(
 \begin{array}{c}
      \langle \lambda_1 |_Q     \\
      \langle \lambda_2 |_Q     \\
      \vdots 
 \end{array}
\right) 
\end{equation}
by using the diagonalizing operator $P$ of $H$, which has a structure as 
$P=( |\lambda_1 \rangle, |\lambda_2 \rangle, \protect\linebreak\ldots  )$, where $| \lambda_i \rangle$ are 
eigenstates of $H$. 
We note that $``P^{\dag_Q}"$ is defined by using the $Q$-hermitian conjugate of kets, 
so $``P^{\dag_Q}" \neq Q^{-1} P^\dag Q$. 
Then we see that $``P^{\dag_Q}" P = {\bold 1}$, namely, $``P^{\dag_Q}"=P^{-1} $. 
Hence we can say that $P$ is $Q$-unitary.

Next we consider the relation $``P^{\dag_Q}" H P = D $. 
The $(i,j)$-component of this relation in $|\lambda_i  \rangle$ basis is written as 
$\langle \lambda_i |_Q H | \lambda_j \rangle = \lambda_i \delta_{ij} $. 
Taking the complex conjugate, we obtain $\langle \lambda_j |_Q H^{\dag_Q} | \lambda_i \rangle = \lambda_i^* \delta_{ij} $, 
that is to say, $\langle \lambda_i |_Q H^{\dag_Q} | \lambda_j \rangle = \lambda_i^* \delta_{ij} $. 
This is written in the operator form as $``P^{\dag_Q}" H^{\dag_Q} P = D^\dag $. 
Therefore we obtain 
\begin{equation}
[H, H^{\dag_Q} ] = P [D, D^\dag ] P^{-1} =0.
\end{equation} 
Thus $H$ is $Q$-normal. 
In other words the inner product $I_Q$ is defined so that 
$H$ is normal with regard to it.

Furthermore for later convenience we decompose $H$ as $H=H_{Qh} + H_{Qa}$, 
where $H_{Qh} = \frac{H + H^{\dag_Q} }{2}$ and $H_{Qa} = \frac{H - H^{\dag_Q} }{2}$ 
are $Q$-hermitian and anti-$Q$-hermitian parts of $H$ respectively. 
If we decompose $D$ as $D=D_R + iD_I$, where 
the diagonal components of $D_R$ and $D_I$ are the real and imaginary parts of 
the diagonal components of $D$ respectively, 
$H_{Qh} $ and $H_{Qa}$ can be expressed as $H_{Qh}=P D_R P^{-1}$ and $H_{Qa}=i P D_I P^{-1}$.

\vspace*{1mm}
\noindent
{\it Normalization of $| \psi \rangle$ and expectation value}\hspace*{7mm}
We consider some state $| \psi(t) \rangle $, which obeys 
the Schr\"{o}dinger equation 
$i \hbar \frac{d}{dt} | \psi (t) \rangle = H | \psi (t) \rangle $. 
Normalizing it as 
\begin{equation}
|\psi(t) \rangle_{N} \equiv \frac{1}{\sqrt{ \langle {\psi}(t) |_Q ~{\psi}(t) \rangle} } | {\psi}(t) \rangle ,
\end{equation} 
we define the expectation value of some operator ${\cal O}$ by 
\begin{equation}
\bar{\cal O}_Q(t) \equiv  {}_{N} \langle \psi(t) |_Q {\cal O} | \psi(t) \rangle_{N} 
=  {}_{N} \langle \psi(t_0) |_Q {\cal O}_{QH}(t-t_0) | \psi(t_0) \rangle_{N},
\end{equation} 
where we have introduced the time-dependent operator in the Heisenberg picture, 
${\cal O}_{QH}(t-t_0) \equiv  \frac{ \langle \psi(t_0) |_Q \psi(t_0) \rangle }{ \langle \psi(t) |_Q \psi(t) \rangle } 
e^{ \frac{i}{\hbar} H^{\dag_Q} (t-t_0) } {\cal O} e^{ -\frac{i}{\hbar} H(t-t_0) }$. 
Since the normalization factor depends on time $t$, $| \psi(t) \rangle_{N} $ 
obeys the slightly modified Schr\"{o}dinger equation, 
\begin{equation}
i\hbar \frac{d}{dt} | \psi(t) \rangle_{N} 
= H_{Qh} | \psi(t) \rangle_{N} 
+ \left( H_{Qa} -{}_{N} \langle \psi(t) |_Q H_{Qa} | \psi(t) \rangle_{N} \right) | \psi(t) \rangle_{N}. \label{sch}
\end{equation}
In addition ${\cal O}_{QH}$ obeys the slightly modified Heisenberg equation, 
\begin{equation}
i\hbar \frac{d}{dt} {\cal O}_{QH} 
= [ {\cal O}_{QH}, H_{Qh} ] 
+ \left\{  {\cal O}_{QH} , H_{Qa} -  {}_{N} \langle \psi(t) |_Q H_{Qa} | \psi(t) \rangle_{N} \right\} . \label{hei}
\end{equation}
In eqs.(\ref{sch})(\ref{hei}) we find the effect of $H_{Qa}$, the anti-$Q$-hermitian part of the Hamiltonian $H$, 
though it seems to disappear in the classical limit. 
But with the second procedure we explain next, 
we shall see that 
the effect of $H_{Qa}$ disappears.

\vspace*{1mm}
\noindent
{\it The mechanism for suppressing the anti-$Q$-hermitian part of the Hamiltonian}\hspace*{7mm}
To show the mechanism for suppressing the effect of $H_{Qa}$, 
we shall see the time development of $| \psi(t) \rangle$ explicitly. 
We introduce $| \psi'(t) \rangle $ by 
$| \psi' (t) \rangle = P^{-1} | \psi (t) \rangle$, and expand it as 
$| \psi'(t) \rangle = \sum_i a_i(t) | e_i \rangle $. 
Then $| \psi(t) \rangle $ can be written in an expanded form as 
$| \psi(t) \rangle = \sum_i a_i(t) | \lambda_i \rangle $. 
Since $ | \psi'(t) \rangle$ obeys 
$i \hbar \frac{d}{dt} | \psi'(t) \rangle =D | \psi'(t) \rangle$, 
the time development of $| \psi(t) \rangle$ from some time $t_0$ is calculated as 
\begin{eqnarray}
| \psi(t) \rangle &=& P e^{- \frac{i}{\hbar} D (t-t_0)} | \psi'(t_0) \rangle \nonumber\\
&=& \sum_i a_i(t_0) 
e^{ \frac{1}{\hbar} \left( \text{Im} \lambda_i - i \text{Re} \lambda_i \right) (t-t_0)}       
| \lambda_i \rangle . \
\end{eqnarray}

$\text{Im} \lambda_i $  corresponds to the anti-$Q$-hermitian part of the Hamiltonian 
since $H_{Qa} = i P D_I P^{-1}$. 
As for the anti-$Q$-hermitian part $H_{Qa}$, 
we can crudely imagine that 
some of $\text{Im} \lambda_i $ take the maximum value $B$, 
say, we assume that $H$ is bounded.
We denote the corresponding subset of $\{ i \}$ as $A$. 
Then we can Taylor-expand $H_{Qa}$ around its maximum 
and get a good approximation to the practical outcome of the model. 
In the Taylor-expansion we do not have the linear term because we expand it near the maximum, 
so we get only non-trivial terms of {\em second order}. 
In this way $H_{Qa}$ becomes constant in the first approximation, 
and it is not so important observationally.
Therefore, if a long time has passed, namely for large $t-t_0$, 
the states with $\text{Im} \lambda_i |_{i \in A}$ survive and contribute most in the sum.

To show how $| \psi(t) \rangle $ is effectively described for large $t-t_0$, 
we introduce a diagonalized Hamiltonian $\tilde{D}_{R}$ as 
\begin{equation}
\langle e_i | \tilde{D}_{R} | e_j \rangle \equiv 
\left\{ 
 \begin{array}{cc}
      \langle e_i | D_R | e_j \rangle =\delta_{ij} \text{Re} \lambda_i  & \text{for} \quad i \in A , \\
      0 &\text{for} \quad i \not\in A , \\ 
 \end{array}
\right. \label{DRtilder}
\end{equation}
and define $H_{\text{eff}}$ by $H_{\text{eff}} \equiv P \tilde{D}_{R} P^{-1}$. 
$H_{\text{eff}}$ is $Q$-hermitian, $H_{\text{eff}} ^{\dag_Q} =H_{\text{eff}}$, 
and satisfies 
$H_{\text{eff}} | \lambda_i \rangle = \text{Re}\lambda_i | \lambda_i \rangle$.  
Furthermore, we introduce 
$| \tilde\psi(t) \rangle \equiv \sum_{i \in A}  a_i(t) | \lambda_i \rangle $. 
Then $| \psi(t) \rangle$ is approximately estimated as 
\begin{eqnarray}
| \psi(t) \rangle 
&\simeq& e^{ \frac{1}{\hbar} B (t-t_0)} 
\sum_{i \in A}  a_i(t_0) e^{-\frac{i}{\hbar} {\text Re} \lambda_i (t-t_0)} | \lambda_i \rangle \nonumber\\
&=&e^{ \frac{1}{\hbar} B (t-t_0)}  e^{-\frac{i}{\hbar} H_{\text{eff}} (t-t_0)} | \tilde\psi(t_0) \rangle 
= | \tilde\psi(t) \rangle . \label{psiprimetket}
\end{eqnarray}
The factor $e^{ \frac{1}{\hbar} B (t-t_0)} $ in Eq.~(\ref{psiprimetket}) 
can be dropped out by normalization. 
Thus we have effectively obtained a $Q$-hermitian Hamiltonian $H_{\text{eff}}$ 
after a long time development though our theory is described 
by the non-hermitian Hamiltonian $H$ at first. 
Indeed the normalized state 
\begin{equation}
| \psi(t) \rangle_{N} 
\simeq \frac{1}{\sqrt{ \langle \tilde{\psi}(t) |_Q ~\tilde{\psi}(t) \rangle} } | \tilde{\psi}(t) \rangle \nonumber \equiv | \tilde{\psi}(t) \rangle_{N} 
\end{equation}
time-develops as 
$| \tilde{\psi}(t) \rangle_{N} 
=e^{-\frac{i}{\hbar} H_{\text{eff}} (t-t_0)} | \tilde{\psi}(t_0) \rangle_{N}$.
We see that the time dependence of the normalization factor 
has disappeared due to the $Q$-hermiticity of $H_{\text{eff}}$. 
Thus $| \tilde{\psi}(t) \rangle_{N}$, the normalized state 
by using the inner product $I_Q$, obeys the Schr\"{o}dinger equation  
\begin{equation}
i\hbar \frac{\partial}{ \partial t} | \tilde\psi(t) \rangle_{N} = H_{\text{eff}} | \tilde\psi(t) \rangle_{N}.
\end{equation}
On the other hand, the expectation value is given by 
\begin{equation}
\bar{\cal O}_Q(t) \simeq  {}_{N} \langle \tilde\psi(t) |_Q {\cal O} | \tilde\psi(t) 
\rangle_{N} 
= {}_{N} \langle \tilde\psi(t_0) |_Q \tilde{\cal O}_{QH}(t-t_0) | \tilde\psi(t_0) \rangle_{N},
\end{equation} 
where we have defined a time-dependent operator $\tilde{\cal O}_{QH}$ 
in the Heisenberg picture by 
$\tilde{\cal O}_{QH}(t-t_0) \equiv e^{ \frac{i}{\hbar} H_{\text{eff}} (t-t_0) } {\cal O}  e^{ -\frac{i}{\hbar} H_{\text{eff}}(t-t_0) }$. 
We see that $\tilde{\cal O}_{QH}$ obeys the Heisenberg equation 
\begin{equation}
\frac{d}{dt} \tilde{\cal O}_{QH}(t-t_0) = \frac{i}{\hbar} [ H_{\text{eff}}, \tilde{\cal O}_{QH} (t-t_0)].
\end{equation}

As we have seen above, the non-hermitian Hamiltonian $H$ has become 
a hermitian one $H_\text{eff}$ automatically 
with the proper inner product $I_Q$ and the mechanism of suppressing the 
anti-hermitian part of $H$ after a long time development. 
If $H$ is written in a local form like $H= \frac{1}{2m}p^2 + V(q)$, 
does the locality remain even after $H$ becomes hermitian?
It is not clear, but for the moment  
let us assume that the hermitian Hamiltonian $H_\text{eff}$ has a local expression like 
$H_{\text{eff}} \simeq \frac{1}{2m_{\text{eff}}} p_{\text{eff}}^2 
+ V_{\text{eff}}(q_{\text{eff}})$, 
and see probability conservation. 
Besides the usual $q_{\text{eff}}$-representation of the state $|\tilde{\psi}(t) \rangle_N$, 
$\tilde{\psi}(q_{\text{eff}}) \equiv \langle q_{\text{eff}} | \tilde{\psi}(t) \rangle_N$, 
we introduce $\tilde{\psi}_Q (q_{\text{eff}}) \equiv \langle q_{\text{eff}} |_Q \tilde{\psi}(t) \rangle_N$, 
and define a probability density by 
\begin{equation}
\rho_{\text{eff}}=\tilde{\psi}_Q(q_{\text{eff}})^{*} \tilde{\psi}(q_{\text{eff}})
={}_N\langle \tilde{\psi}(t) |_Q q_{\text{eff}} \rangle 
\langle q_{\text{eff}} | \tilde{\psi}(t) \rangle_N. 
\end{equation}
Then, since we have $i\hbar \frac{\partial}{\partial t} \tilde{\psi}(q_{\text{eff}}) = H_{\text{eff}} \tilde{\psi}(q_{\text{eff}})$ 
and $i\hbar \frac{\partial}{\partial t} \tilde{\psi}_Q(q_{\text{eff}}) = H_{\text{eff}}^* \tilde{\psi}_Q(q_{\text{eff}})$, 
we obtain a continuity equation 
\begin{equation}
\frac{\partial \rho_{\text{eff}}}{\partial t} 
+ \frac{\partial}{\partial q_{\text{eff}}} j_{\text{eff}}(q_{\text{eff}},t) = 0,
\end{equation} 
where $j_{\text{eff}}(q_{\text{eff}},t)$ is a probability current density defined by 
\begin{equation}
j_{\text{eff}}(q_{\text{eff}},t)= \frac{i\hbar}{2 m_{\text{eff}}} 
\left( \frac{\partial}{\partial q_{\text{eff}}} \tilde{\psi}_Q^* \tilde{\psi}       
- \tilde{\psi}_Q^* \frac{\partial}{\partial q_{\text{eff}}} \tilde{\psi} \right).
\end{equation} 
Thus we see that if $H_{\text{eff}}$ has a local expression, we have the probability conservation 
$\frac{d}{dt} \int \rho_{\text{eff}} dq_{\text{eff}}=0$.

\vspace*{1mm}
\noindent
{\it Discussion}\hspace*{7mm}
In this paper we have studied a system described by a non-hermitian diagonalizable Hamiltonian $H$. 
For a measurement to be physically reasonable, 
we have introduced the proper inner product $I_Q$ so that $H$ gets normal 
with regard to it, and defined $Q$-hermiticity, i.e. hermiticity with regard to $I_Q$. 
Next we have explicitly presented the mechanism for suppressing the effect of 
the anti-$Q$-hermitian part of $H$ after a long time development, 
and thus effectively obtained the hermitian Hamiltonian $H_{\text{eff}}$. 
This result suggests that we have no reason to 
maintain that at the fundamental level the Hamiltonian should be hermitian. 
Furthermore we have seen that if $H_{\text{eff}}$ is written in a local form, 
we obtain the continuity equation leading to probability conservation.

Finally let us discuss an estimation of a state at an early time $t_1$. 
It is expressed as 
$| \psi_\text{true}(t_1) \rangle_N =  e^{ - \frac{i}{\hbar} H (t_1 -t_0)} 
\sqrt{ \frac{ \langle \psi(t_0) |_Q \psi(t_0) \rangle }{ \langle \psi(t_1) |_Q \psi(t_1) \rangle } } | \psi(t_0) \rangle_N$. 
%
But if a historian who lives at a late time $t$ were asked about the state at $t_1$, 
he would extrapolate back in time from his own time $t$ 
by using the phenomenological Hamiltonian $H_{\text{eff}}$ rather than 
the fundamental one $H$, because at the late time he would only know the hermitian Hamiltonian.  
Thus he would specify the early state at time $t_1$ as
%
$|\psi_\text{historian}(t_1)\rangle_N 
= e^{- \frac{i}{\hbar} H_{\text{eff}}(t_1 -t)} e^{ -\frac{i}{\hbar} H (t -t_0)}
\sqrt{ \frac{ \langle \psi(t_0) |_Q \psi(t_0) \rangle }{ \langle \psi(t) |_Q \psi(t) \rangle } } |\psi(t_0) \rangle_N$. 
%
This is a false picture and different from the true state 
$| \psi_\text{true}(t_1) \rangle_N $ because $H_{\text{eff}} \neq H$. 
Actually, the seeming past state 
$| \psi_\text{historian}(t_1) \rangle_N$ 
will be mainly a superposition of the eigenstates correlated to the subset $A$. 
Since the set of the eigenstates correlated to the subset $A$  
is much smaller than that of all the eigenstates of $H$, 
the seeming state 
$| \psi_\text{historian}(t_1) \rangle_N$ would look like necessarily having come from 
a special rather tiny part of the full Hilbert space in the fundamental theory. 
In other words, it would look to the historian that the universe necessarily had begun in a state 
inside the rather tiny subspace of the fundamental Hilbert space with the highest imaginary part of the eigenvalues of the Hamiltonian. 
Thus the fundamentally true initial state $| \psi(t_0) \rangle_N$ 
tends to be hidden from the historian at the late time more and more as the 
time $t$ gets later and later.  
This story implies that 
if our universe had begun with a non-hermitian Hamiltonian at first in some fundamental theory, 
then we could misestimate the early state at the time $t_1$ by using the hermitian 
Hamiltonian to extrapolate back in time.

We note that the above story is based on the assumption that the correspondence principle
between a quantum regime and a classical one holds in our system, which 
we have not explicitly studied in this paper. 
At the point where the imaginary part of the action $S_I$ is minimized,
we have $\delta S_I \simeq 0$, so that in the region around it
$S_I$ is constant practically.
Thus we see little effect of $S_I$ there.
This is consistent with our observation that the anti-hermitian part
of the Hamiltonian is suppressed after a long time, 
but it is not a trivial problem to examine the classical behavior explicitly. 
We postpone this problem to a future study.


\vspace*{3mm}

This work was supported by Danish Natural Science Research Council (FNU, Denmark) 
and Grants-in-Aid for Scientific Research (Nos~18740127 and 21740157) 
from the Ministry of Education, Culture, Sports, Science and Technology (MEXT, Japan), 
and begun during the stay of one of the authors (K.N.) at NBI by 
using the JSPS Researcher Exchange Program FY2008. 
K.N. would like to thank Jan Ambjorn, who accepted his stay initially, 
Klara Pavicic and all the other members and visitors of NBI for their kind hospitality.

\end{document}